\documentstyle[11pt,epsfig]{article}
\def\baselinestretch{1.3}
\newcommand{\ba}{\begin{array}}
\newcommand{\ea}{\end{array}}
\newcommand{\bd}{\begin{displaymath}}
\newcommand{\ed}{\end{displaymath}}
\newcommand{\be}{\begin{equation}}
\newcommand{\ee}{\end{equation}}
\newcommand{\bea}{\begin{eqnarray}}
\newcommand{\eea}{\end{eqnarray}}

\def\q2 {q^2}

\def\lsim{\mathrel{\raise.3ex\hbox{$<$\kern-.75em\lower1ex\hbox{$\sim$}}}}
\def\gsim{\mathrel{\raise.3ex\hbox{$>$\kern-.75em\lower1ex\hbox{$\sim$}}}}

\begin{document}

\begin{center}
{\Large\bf Split supersymmetry and the role of a  
light fermion\\ in a supergravity-based scenario}\\[20mm]
Biswarup Mukhopadhyaya\footnote{E-mail: biswarup@mri.ernet.in}\\
{\em Harish-Chandra Research Institute,\\
Chhatnag Road, Jhusi, Allahabad - 211 019, India}\\
Soumitra SenGupta \footnote{E-mail: tpssg@iacs.res.in} \\
{\em Departmernt of Theoretical Physics,\\
Indian Association for the Cultivation of Science\\
Jadavpur, Kolkata- 700 032, India}
\\[20mm]
\end{center}
\begin{abstract}
We investigate split supersymmetry (SUSY)  
within a supergravity framework, where  local SUSY is broken by the F-term
of a hidden sector chiral superfield X. With reasonably general assumptions, 
we show that the fermionic component of X will always have mass within 
a Tev. Though its coupling to the observable sector superfields is 
highly suppressed in Tev scale SUSY, we show that it can be enhanced 
by many orders in split SUSY, leading to its likely participation in 
accelerator phenomenology.We conclude with a specific example of such a scenario
in a string based supergravity model.

\end{abstract}
\vskip 1 true cm
\newpage
\setcounter{footnote}{0}
\def\baselinestretch{1.8}

Supersymmetry (SUSY), that is to say, a symmetry between fermions 
and bosons, has long been
deemed to be phenomenologically significant if it is broken within the
TeV scale, whereby the quadratic divergence in the Higgs boson mass
is tamed, and the superpartners of all the standard model (SM) particles 
lie within this mass range. This age-old dictum has been given a jolt by the 
recently suggested notion of split supersymmetry \cite{arkani1}
, which allows the SUSY breaking 
scale, and all scalars except one light Higgs to be much higher than a TeV,
while at the same time retaining light fermionic superparticles. Although this
destroys the cancellation of quadratic divergence, it is argued that having
a fine-tuned Higgs mass is not that unexpected, since one has to resort to
more severe fine-tuning anyway in order to suppress the cosmological constant
in a broken SUSY. Such a possibility is inspired by the landscape scenario
in string theories, which opens up the possibility of a large multitude 
of vacua \cite{landscape,susskind,douglas,kachru,dine1,hamed}.
Living in a universe with unexpectedly small Higgs mass and cosmological
constant, it is argued, is quite possible for us if we happen to be located
around one such vacuum which supports galactic structures \cite{structure,vilenkin,rees}. 
Once the 
Higgs mass is not its responsibility, SUSY, possibly an 
artifact of superstrings,
can be broken at an arbitrary high energy scale. However, it not only
suppresses flavour-changing neutral current (FCNC) and proton decay,
but also helps the unification of gauge coupling \cite{uni}
and  supplies a cold dark matter candidate  in 
the acceptable range \cite{dm0,dm,dm1}
if the gauginos and Higgsinos are within a TeV. Thus one is now faced
with a new possibility-- namely, only the SM spectrum (including 
one Higgs scalar)
plus gauginos and Higgsinos within the search limit of particle accelerators,
but all other scalars way upward. There is no flavour-changing neutral 
current problem here due to decoupling. Furthermore, the undesirable
possibility of gluinos having as large a lifetime as the age of the universe
restricts the SUSY breaking scale to be $\sim10^{13}$ GeV \cite{arkani1}. 
Phenomenological consequences 
of split SUSY \cite{pheno,pheno1,pheno2,pheno3,pheno4,pheno5,pheno6,pheno7,pheno8,pheno9,pheno10,pheno11,pheno12,pheno13} as well as 
various aspects of theoretical models that can lead to 
it \cite{theo1,theo1b,theo1c,theo1d,theo1e,theo1f,theo1g,theo1h,theo1i,theo1j,theo1k,theo1l,theo1m,theo1n,theo1o,theo1p,theo1q,theo1r}
have been widely discussed in recent times.

In this note we look at some general consequences of SUSY breaking within
an N=1 supergravity (SUGRA) scenario. A hidden sector chiral superfield X 
is assumed responsible for all soft SUSY breaking terms in the 
observable sector through its nonrenormalizable interaction with the superfields 
in the minimal SUSY standard model (MSSM).

We make no specific model assumption excepting the postulate that SUSY is broken
in the hidden sector via F-terms. In addition, 
we include, in a phenomenological approach, the possibility of having
a large mass hierarchy between 
the scalars and the fermions, which can give rise to split SUSY. 
In order to achieve this, we assume that various terms involving the
chiral superfield X, arising from both the Kahler potential and the
superpotential, can have multiplicative factors that may in principle
arise from other new physics effects. Such multiplicative
factors are treated as free parameters, capable of being very small.  
It is then shown that if all soft terms as well as the Higgsino
mass term in the superpotential originate from interactions from X, then
the requirement of gauginos and Higgsinos within a TeV implies that
the fermionic component of X also must be within this range. Moreover, while
this light fermion (called $\psi_X$ here) has extremely small coupling
with the MSSM fields in TeV scale SUSY, the strength of such coupling can be
much higher if the SUSY breaking scale is high. Thus one has to admit the 
possibility of low-energy phenomenology being altered by the field  $\psi_X$
(which can even turn out to be a dark matter component).
This is a reaffirmation of our earlier conclusion \cite{akuami} in the special
context of a braneworld model of split SUSY. However, it is now established
on more general grounds.

The most general action for a supergravity multiplet coupled to 
chiral superfield and super Yang-Mill field is given by \cite{sugra1,sugra1a,sugra1b,sugra1c}
\begin{equation}
\int d^4x d^4\theta {K(S,\bar{S} e^{2V})} + \int{d^4 x} Re \int{d^2 
\theta W(S)} + 
\int {d^4x} Re \int {d^2\theta {f_{\alpha\beta}(S)W^{\alpha}W^{\beta}}}
\end{equation}
where $K(S,\bar{S})$ is an arbitrary real function of the chiral multiplets 
$S_i$ and and their conjugates, $W(S)$ is a holomorphic function of  
$S_i$, V is the vector multiplet of a Yang-Mill gauge
group $U$ and $W^{\alpha}$ is the corresponding field strength.
The analytic gauge kinetic function $f$ determines the kinetic 
terms for the fields in the vector
multiplet $V$ and the gauge coupling constant $Re {f_a} = 1/g_a^2$, where 
the index {\it a} is associated with different
gauge groups $U_a$ with total gauge group $U= \Pi { U_a}$.\\ 
If now one defines a function
\begin{equation}
G = K + log|W|^2
\end{equation}
then all mass terms and couplings are determined by the vacuum expectation 
values (vev) of
appropriate functions of $G$, it's derivatives and the chiral scalars.
As with global supersymmetry, spontaneous breakdown of local supersymmetry 
may take place either as $F$ term breaking and/or a $D$ term breaking. In 
either case one of the auxiliary component of
superfield must develop a nonzero vev. In this work we focus on a 
$F$-type breaking and the
relevant term under consideration is,   
\begin{equation}
F_i = M_P e^{G/2M_P^2}(G^{-1})^j_i G_j + f^*_{,k}
(G^{-1})^k_i \bar{\lambda}\lambda + .......
\end{equation}
where $\lambda$ is a gaugino field.
We further concentrate on scenarios where the local SUSY is broken in 
the hidden sector by the usual mechanisms \cite{sugra2}
under which the auxiliary component of a chiral supermultiplet acquires a non-zero
vacuum expectation value (vev) or gaugino condensation takes place in the 
hidden sector gauge group at some condensation scale where the gauge 
theory becomes strongly coupled. In either case, a superpotential
$W(X)$ ( where X is either some hidden sector chiral scalar or an 
effective scalar originated from the gaugino bilinear )
develops in the hidden sector resulting into breakdown of local SUSY at 
some scale, say vev of the scalar component of the 
chiral superfield or condensation scale. This breaking is mediated to the 
observable sector through gravitational
interaction leading to the generation of soft SUSY breaking parameters 
in the observable sector. If $C^{\alpha}$'s are generic observable sector 
fields then the full Kahler potential $K$ and the superpotential
$W$ can be expanded ( suppressing the dummy indices between $C$'s and it's coefficients) 
around the hidden sector field X as
\begin{eqnarray}
K = K_1(X,\bar{X}) + K_2(X,\bar{X})\bar{C}C + K_3(X,\bar{X})CC + ............\\
W = W_1(X) + W_2(X)CC + W_3(X)CCC + ............
\end{eqnarray}
Various masses of the observable sector fields and their couplings can be 
determined from these 
coefficients $K_i$'s and $W_i$'s.
Different terms in the SUSY Lagrangian generate
masses for the sfermions, gauginos, Higgsinos as well as the scalar and
spinor components of X. We
now describe the exact expressions for the different mass terms
that arise from the corresponding terms in the Lagrangian. 

First, we note that the generic  scalar mass arises from the term
\begin{equation}
\int d^4 \theta X^{\dagger}X Q^{\dagger}Q 
\end{equation}
and is given by,
\begin{equation} 
m_0  = F_X/M_P 
\end{equation}

\noindent
If we wish the SUSY breaking scale to be $\simeq 10^n~GeV$,
then we have

\begin{equation} 
F_X/M_P = 10^n
\end{equation}
or, in other words, $F_{X}~\simeq~10^{19+n}~GeV^2$.

The observable sector gaugino mass $M_{1/2}$  (assuming it to be universal)   
originates from (dropping coefficients of the different terms)
\begin{equation}
\int d^2 \theta  f X W_{\beta} W^{\beta} + h.c.+ 
\int d^4 \theta  X^{\dagger} X W_{\beta} W^{\beta}
\end{equation}
where W can be expressed in terms of the components of a gauge supermultiplet
\begin{equation}
W_{\alpha}~=~4i\lambda_{\alpha} - \left[4 \delta^{\beta}_{\alpha} D
+ 2i (\sigma^{\mu} \bar{\sigma}^{\nu})^{\beta}_{\alpha} V_{\mu\nu}\right]
\theta_{\beta} + 4 \theta^2 \sigma^{\mu}_{{\alpha}{\dot{\alpha}}}
\partial_{\mu}\bar{\lambda}^{\dot{\alpha}}
\end{equation}
\noindent
D and $V_{\mu\nu}$ being respectively the auxiliary part of
the vector superfield and the field strength of the gauge boson
$V_\mu$.\\

The term arising from the superpotential in the above expression 
is subject to a suppression
factor $f$ in order to allow the possibility of gaugino masses being
smaller than the scalar masses (or the observable sector SUSY breaking scale) 
by the same factor. In order to achieve split SUSY, the factor f will
be as small as the level of `splitting' required, caused by instruments of
specific models, such as R-symmetry or  gaugino condensation.

Thus the contributions to gaugino mass turn out to be
\begin{equation}
m_{1/2} = fF_X/M_P +  F_X^2/M_P^3 
\end{equation}

Of course, this corresponds to the scenario of a universal gaugino mass
at the SUSY breaking scale. Gaugino non-universality will require one to
assume something like a non-minimal Kahler potential, whereby the second term 
in (9) can be different for different gauginos.

If now $m_{1/2}$ is required to be within a TeV, each of the terms on the
right-hand side in the above equation must not exceed that value, since they
arise from different sources and an accidental cancellation between them is 
unlikely. Using the value of $F_X$, this immediately disallows $n\gsim 11$, 
thus imposing a rather strong constraint on the split SUSY scale. At the same 
time, the restriction on the magnitude of the 
first term on the right-hand side leads to 

\begin{equation}
f\lsim 10^{-8}~for~ n=11;~f = 10^{3-n}~for~n\lsim 11
\end{equation}

Thus the generic SUGRA scenario, with hidden sector F-term breaking,
does not allow scalar masses above about $10^{11}$ GeV.

Before proceeding further, let us note that masses of both the scalar and 
fermion components of $x$ arise from the term

\begin{equation}
\int d^4 \theta (X^{\dagger}X)^2
\end{equation}
arising from the Kahler potential, giving

\begin{equation}
m_{X} = F_X/M_P
\end{equation}
and
\begin{equation}
m_{\psi_X} = XF_X/M_P^2
\end{equation}

In the most general case, the following terms in the SUGRA Lagrangian 
can lead to the term $\mu H_u H_d$ in the MSSM superpotential

\begin{equation}
\int d^2 \theta f_1 X H_u H_d  ~+~  \int d^4 \theta  f_2 X^{\dagger} H_u H_d  
~+~ \int d^4 \theta X^{\dagger}X H_u H_d  
\end{equation}
where the suppression factors $f_1$,  $f_2$ are model-dependent; various 
things, ranging from nonperturbative effects to Peccei-Quinn symmetries, 
have been invoked \cite{witten}, \cite{mu1}, \cite{mu2},  \cite{mu3,mu3a,mu3b} 
to justify their smallness. In fact, similar effects have been claimed to arise
in braneworld scenarios such as that described in \cite{arkani1}, playing roles 
analogous to those of the factors $f$, $f_1$ and $f_2$ in generating various terms.   
Even without making any specific model assumption,
one can, however, see that the Higgsino mass $\mu \simeq 1~TeV$ requires

\begin{equation}
\mu = f_{1} X  +   f_{2}F_X/M_P    +  XF_X/M_P^2 \simeq 10^3
\end{equation}

\noindent
Again, the requirement that each term should not exceed the stipulated value of 
$\mu$ means

\begin{equation}
XF_X/M_P^2 \lsim 10^3
\end{equation}
which immediately implies $m_{\psi_X} \lsim 10^3$. 

Whether such a light fermion will be phenomenologically significant
depends on the strength of its interaction with other fields in MSSM.
Take, for example, the first term in the expression for $\mu$, where
$f_1$ is the strength of $\psi_{X} - \tilde{H}_u - H_d$ interactions.
A further constraint on $f_1$  follows from the observation that both the 
first and third terms contributing to $\mu$  can also give rise to the
bilinear soft breaking term $B\mu$:

\begin{equation}
B\mu = f_1 F_X + F_X^2/M_P^2  
\end{equation}

The contribution from the second term is on the same order as 
$m_0^2$, the square of the SUSY breaking scale, as expected. We have
to demand further that 

\begin{equation}
f_1 F_X \lsim 10^{2n}  
\end{equation}
or, inserting the value of $F_X$,
\begin{equation}
f_1  \lsim 10^{n-19}  
\end{equation}

Thus we are led to interesting conclusions. First, by following this
approach, one does not end up with $B\mu ~\simeq~TeV^2$, which 
would have caused  a problem with the value of $\tan \beta$ \cite{theo1l}.
Secondly, and more crucially, although one has a light fermion
$\psi_X$, it has no perceptible effect on phenomenology when the SUSY
breaking scale in the observable sector is $\simeq$ TeV, because of the
stringent upper limit on its interaction with the Higgs-Higgsino sector.
However, as n increases, or as the SUSY scenario is more and more `split',
the interactions strength also increases, and it can be as large as 
$10^{-8}$ for $n = 11$. As we have shown in an earlier work \cite{akuami}, 
such a strength
of the $\psi_{X} - \tilde{H}_u - H_d$ coupling can be envisioned in
a braneworld scenario leading to split SUSY, and can be phenomenologically
significant. For example, for $m_{1/2} > \mu >m_{\psi_{X}}$, the Higgsino-type
lightest neutralino can then decay into  $\psi_{X}$ and the Higgs 
in about $10^{-8}$ seconds, thus having $\psi_X$ as the invisible superparticle
controlling the final state in SUSY cascades in collider experiments.

Another point to note here is that although $\psi_{X}$ can be light here, 
its coupling to the gauge-gaugino pair is always a dimension-five operator,
suppressed by the Planck mass. Therefore, such coupling does not upset
the phenomenology of Tev-scale MSSM, even if the 
suppression factor $f$ present in 
such an interaction term be of order unity.

Let us re-iterate that our claim that the various suppression factors can be
accommodated within the framework is somewhat phenomenological.
Our purpose is just to emphasize that if there is F-term SUSY breaking and split
SUSY has to be nonetheless a viable possibility, then there must be room
for such suppression. It may be noted that the gaugino mass may acquire a correction
from anomaly mediation and this correction depends on the gravitino mass.  Since
the anomaly-mediated contribution to the {\it i}th gaugino mass is
$\sim g^2_i m_{3/2}/(16\pi^2)$ (where $m_{3/2}$ is the gravitino mass and 
$g_i$ is the corresponding gauge coupling) \cite{amsb}, 
this contribution remains within the order of a TeV if the gravitino mass
is within apppximately $10^5$ GeV. As has been shown in reference \cite{dm0},
such a gravitino is viable cosmologically and otherwise, so long as the lightest
neutralino mass is within about 100 GeV. In addition, it is possible to have
the anomaly contribution  within control for even higher values of $m_{3/2}$,
with specific theoretical assumptions.
 
Thus one can have a split SUSY spectrum well within a supergravity 
framework.  With the gravitino mass given by $m_{3/2} \sim M_{P}e^{K(X)}W(X)$,
an appropriate choice of
the Kahler potential $K(X)$ and superpotential $W(X)$ may 
lead to $m_{3/2}$ within the ranges specified above, 
while SUSY breaking scale is still on the order of $10^{n}$ Gev 
In such a situation one can observe the  
consequences of a light $\psi$ in the most pronounced form, with  
$f_1$ at its upper limit.

It may be noted from the first term in the left-hand side of 
equation (17) that vev of $X$ must be less than equal to $10^{22-n}$.
So larger is the SUSY breaking scale i.e $n$, smaller is the 
the vev of the hidden sector scalar $X$. In other words ,when the SUSY is maximally split the
hidden sector scalar $X$, responsible for the breakdown of SUSY,has the least vev.
Thus for $n=11$, vev of $X$ (given by $<X>$) is  $\leq 10^{11}$.
However an exact value of the vev of $X$ can only be determined from the knowledge of the
exact form of Kahler and superpotential. If $V$ is the scalar 
potential then, from the requirement of the vanishing cosmological constant 
\begin{eqnarray}
<V> = 0\\ 
\end{eqnarray}
This leads to,
\begin{equation}
<e^G(G_AK^{A}_{\bar{B}}G^{\bar{B}} -3)> = 0
\end{equation}
where the indices A,B runs over various chiral superfields, 
with $G_A$ indicating the
derivative of $G$ with respect to the superfield $\Phi_A$.
But we have already shown that the 
vev of $X$ is also $\leq 10^{11}$. This puts strong  
constraints on the choice of the Kahler and superpotential 
of the hidden sector field $X$. 
It should be noted  that the specific situation 
considered in reference \cite{akuami}, 
based on a braneworld picture, can be considered as a special case of the 
general principle presented here. There, the  appropriate suppression factors 
leading to split SUSY arise  from the hierarchy of a few orders
between the five-and four dimensional Planck masses. However,
they all get mapped into our scheme, giving specific values of $f$,  $f_1$
and $f_2$. On the other hand, the conclusions presented here have not taken 
into account the effects for D-terms or anomaly mediation where a large value
of the scalar mass parameter $m_0$ may cause a split in the spectrum under
appropriate assumptions \cite{arkani1,dm0}.

To illustrate the validity of our result we finally present 
an example of a string-inspired effective 
4-dimensional SUGRA model\cite{ssgpm}.
Consider an $E_8 \times E_8$ heterotic 
string model where the second $E_8$ sector 
corresponds to the gauge group of the hidden sector. Local SUSY 
breaks via gaugino condensation\cite{dine}
in the hidden sector at some condensation scale $\mu$ 
leading to a superpotential contribution from the hidden sector field given by,
\begin{equation}
\Omega(X) = a + b e^{-X/\beta}
\end {equation}
where a and b are constatnts and $\beta$ is the coefficient 
of the $E_8$ gauge coupling beta function or the beta function of
some subgroup of $E_8$, if $E_8$ is broken 
spontaneously during compactification.       
The scalar potential is given by the expression,
\begin{equation}
V = -e^{G}[G_i(G^{-1})^{i}_{j}G^j - 3]
\end{equation}
where
\begin{equation}
G = K(X) + \Phi(y,\omega) + ln|Q(T)W + \Omega(X)|^2
\end{equation}
where 
K(X) and $\Phi(y,\omega)$ are the Kahler potential for the X field,
the axion field T ( y = T + T*) and the
gauge non-singlet scalars $C_i$( $\omega = C_iC^{i*}$), Q(T) is 
the world sheet instanton correction to the
superpotential and W is the usual trilinear superpotential for 
the fields $C_i$.\\
Various forms for both $K(X),\Phi(y,\omega)$ have been determined 
for various compactification schemes using classical string
symmetries. However taking into account the string 
loop corrections which breaks some of the 
classical symmetries , a generic form for  $\Phi(y,\omega)$ 
can be written as,
\begin{equation}
\Phi(y,\omega) = -3ln(y-2\omega) + \alpha(y-2\omega)^n
\end{equation}
After a long but straightforward calculation one
finds that the cosmological constatnt vanishes by the condition,
\begin{equation}
<\Phi'^2> = 3 <\Phi''>
\end{equation}
It is now easy to show that in order to have finite vev for the field $y$ and non-vanishing gravitino mass to break supersymmetry,
the value of $n$ must be $ \leq -2$ . As a specific case we choose $n = -2$. This immediately determines masses and vevs 
for various fields.  Here, the hidden sector gauge group becomes strongly coupled and gives rise to gaugino
condensation. 
If $M_c$ is the condensation scale determined by the vev of the hidden sector gaugino
bilinear then,
we obtain the SUSY breaking scale $M_S$ ,the gravitino mass $m_{3/2}$ and the gaugino mass $m_{1/2}$ as,
\begin{equation}
M_{S}^2 \sim M_c^3/M_P 
\end{equation}
\begin{equation}
m_{3/2} \sim M_c^3/{M_{P}^2} \sim m_{1/2}
\end{equation}
while the scalar mass is given as,
\begin{equation}
m^2 = exp( <K> +3/2)<\Omega>^2 (2\alpha/3)^{-3}
\end{equation}
Choosing  $M_c \sim 10^{13-14}$ we get the desired values for
the SUSY breaking scale, gravitino mass and the gaugino mass . The vev for the y field is given by,
\begin{equation}
<y> \sim M_{S}^2/M_P
\end{equation}
and that for the X field is determined from the condition
\begin{equation}
K_{X}(<X>) \Omega((<X>) = \Omega_{X}(<X>)
\end{equation}
where the suffix $X$ implies derivative with respect to the field $X$.
Clearly for a suitable choice for K(X) ( completely arbitrary so far),
one can generate the desired split SUSY scenario with all the features described 
previously.  It should be remembered though, that $<y>$ of a magnitude
implied by equation (33) means a large radius of compactification. 
Such a large compactification radius pulls down the compactification scale and 
the effective field theory therefore breaks down much above the compactification scale. 
This also may lead to difficulties with Grand Unification, as a power law behaviour
of the gauge couplings above the compactification scale tends to reduce the
lifetime of the proton \cite{pr1,pr1a}. However, one should note that the above case is
just an example of achieving split SUSY, where additional special assumptions may
have to be made \cite{pr2} in order to avoid the problems related to 
Grand Unification and dynamical SUSY breaking.

To conclude, a general scheme of hidden sector F-term SUSY breaking,
keeping the provision of arbitrarily large scalar masses but TeV-scale
fermions, is seen to entail a fermion with mass within a TeV.  The strength
of its coupling, although negligibly small in the case of TeV-scale SUSY,
can be considerably higher for a split SUSY scenario with large scalar masses,
and can even have observable effects on final states in SUSY cascades.

{\bf Acknowledgment:} We thank Ashoke Sen for helpful comments.
BM acknowledges the hospitality of Indian Association
for the Cultivation of Science, Kolkata, where this study was initiated. SSG
thanks Harish-Chandra Research Institute for hospitality during the
concluding part of the project.

\end{document}